\documentclass[aps,pra,reprint,showpacs]{revtex4-1}%
\usepackage{graphicx}
\usepackage{dcolumn}
\usepackage{bm}
\usepackage{amsmath}
\usepackage{amsfonts}
\usepackage{amssymb}
\DeclareMathOperator{\Real}{Re}
\DeclareMathOperator{\Imag}{Im}
\DeclareMathOperator{\Tr}{Tr}
\DeclareMathOperator{\mean}{mean}
\DeclareMathOperator{\var}{var}

\begin{document}
\preprint{APS/123-QED}
\title{Violation of the area law and long range correlations in infinite matrix product states}

\author{Anne E. B. Nielsen$^1$}
\author{Germ\'an Sierra$^2$}
\author{J. Ignacio Cirac$^1$}
\affiliation{${}^1$Max-Planck-Institut f{\"u}r Quantenoptik,
Hans-Kopfermann-Strasse 1, D-85748 Garching, Germany}
\affiliation{${}^2$Instituto de F\'isica Te\'orica, UAM-CSIC, Madrid, Spain}

\date{\today}

\begin{abstract}
We propose to construct a family of states of one-dimensional spin chains by replacing the finite dimensional matrices in matrix product states by infinite dimensional operators constructed from bosonic annihilation and creation operators. The resulting states are demonstrated to violate the area law and to exhibit long range correlations. In addition, we propose an efficient way to prepare the states experimentally, in which the spins interact sequentially with an ancilla system.
\end{abstract}


\maketitle

\section{Introduction}\label{introduction}

Even if the underlying physical model is well-known, it is, in general, difficult to simulate quantum many body systems on a classical computer. The difficulty stems from the fact that the dimension of the involved Hilbert space grows exponentially with the size of the system, which often renders a treatment in the full Hilbert space intractable. This motivates the search for families of states that provide a good approximate description of the solution to a specific set of problems but are specified in terms of a much smaller number of parameters.

Considering a one dimensional chain of spins, one such family of states is the matrix product states (MPS). An MPS (see \cite{mpsreview} for a review) with open boundary conditions is a state of the form
\begin{equation}
|\psi\rangle=\sum_{s_1,\ldots,s_N}\langle u|
A_{s_1}^{(1)}\ldots A_{s_N}^{(N)}|v\rangle|s_1,\ldots,s_N\rangle,
\end{equation}
where $A_{s_i}^{(i)}$ are $D\times D$ matrices, $\langle u|$ is a $1\times D$ vector, $|v\rangle$ is a $D\times1$ vector, and $|s_i\rangle$ is the state of the $i$th spin. The number of parameters required to specify the state scales linearly in the number of spins $N$ and quadratically in the virtual dimension $D$. Matrix product states are particularly suitable to approximate the ground state at zero temperature of Hamiltonians with finite range spin-spin interactions and an energy gap between the ground state and the lowest excited state. This appears because these states fulfil the area law \cite{arealaw,reviewarealaw}, which states that the von Neumann entropy of the reduced density matrix of a subchain is proportional to the number of spins at the boundary, i.e., two in one dimension. For an MPS the entropy of a subchain is limited by the logarithm of $D$ \cite{mpsreview}, and it has been shown that all states fulfilling the area law can be approximated by MPS using a number of parameters that scales polynomially in the number of spins \cite{appmps}.

There are, however, a number of interesting cases, where MPSs either fail to approximate the solution or provide less accurate results. A particularly interesting example of the latter is critical systems, where the entanglement entropy of a subchain consisting of $L$ spins scales as $\ln(L)$ \cite{lnL1,lnL2}. It has been shown that such systems can be approximated by MPSs if the virtual dimension grows polynomially in the size of the system \cite{gsrepresentation}, and the multiscale entanglement renormalization ansatz has also been used to investigate critical systems \cite{MERA1,MERA2}. To obtain states with a higher entropy than MPSs, it has been proposed in \cite{cftimps} to replace the finite dimensional matrices $A_{s_i}^{(i)}$ by infinite dimensional chiral vertex operators constructed from a conformal bosonic field \cite{cft}. The resulting states, which are called infinite matrix product states (iMPS) because $D\rightarrow\infty$, are able to describe critical and noncritical systems on an equal footing.

The chiral vertex operators contain a product of infinitely many normal ordered exponentials of bosonic annihilation and creation operators, where the operators in different exponentials commute because they correspond to different modes. Since each of the infinitely many bosonic modes constitutes an infinite dimensional Hilbert space on its own, it is, in fact, sufficient to include just one of them to obtain an iMPS, and the aim of the present paper is to investigate the properties of the family of states that emerge, when we keep only $M$ of the bosonic modes in the chiral vertex operators. We find that these states contain even more entropy than the critical iMPS in \cite{cftimps} and exhibit long range correlations.

A further motivation for investigating this family of states is that the particular construction of the states naturally suggests a way to prepare the states experimentally in which each of the spins in the chain interacts sequentially with $M$ or $M+1$ ancilla systems, whereafter a conditional measurement is applied to each of the ancillas. The preparation scheme is efficient (for moderate $M$) because the required number of operations per ancilla scales linearly in $N$ and we find that the probability of a successful outcome is approximately of order $1/N$ for each of the conditional measurements. We note that the idea of sequential generation is in line with \cite{sequential,sequentialPRA}, where it is shown that all MPSs can be generated deterministically by letting a chain of spins initially in a product state interact sequentially with an ancilla system with $D$ levels, and that all states prepared in this way are instances of MPSs with virtual dimension $D$. In \cite{expcmps}, it is shown that also continuous matrix product states can be generated deterministically by sequential interaction of a continuous quantum system with an ancilla.

The structure of the article is as follows. In Sec.~\ref{Mmodes}, we provide explicit expressions for the states obtained by keeping only $M$ modes in the vertex operators, and we compute entropies and correlation functions of these states to characterize their properties. In Sec.~\ref{implementations}, we propose and discuss a way to generate the states experimentally. The proposed experimental implementation suggests some natural modifications of the scheme, and the influence of these modifications on the properties of the produced states is investigated in Sec.~\ref{modifications}. Section~\ref{conclusion} concludes the paper, and the appendices provide details on the derivations of the analytical results stated in the main part of the article.

\section{iMPS constructed from one or a few bosonic modes}\label{Mmodes}

\subsection{Wave function}

\begin{figure}
\includegraphics[width=0.9\columnwidth]{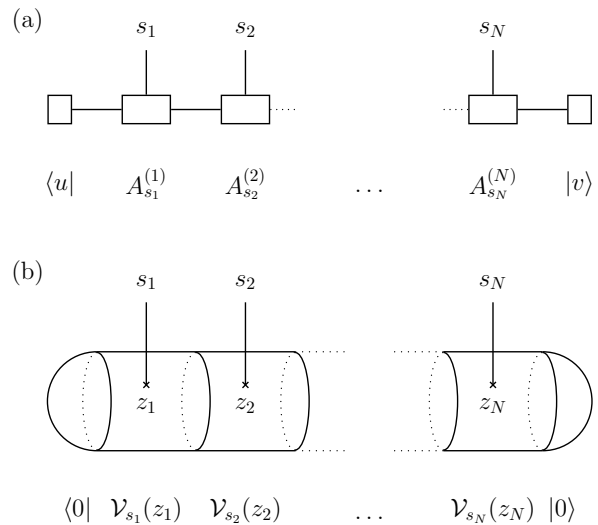}
\caption{(a) Graphical representation of an MPS as a tensor network. The tensors $A_{s_i}^{(i)}$ are drawn as rectangles with two horizontal legs corresponding to the two virtual indices (running from $1$ to $D$) and one vertical leg representing the physical index $s_i$. The input and output states are drawn as rectangles with one leg, and the combination of two legs into a line joining two tensors means contraction of the corresponding indices. (b) In the iMPS construction, the three-legged tensors are replaced by pieces of a cylinder, the vacuum state is represented by a half sphere, and the combination of two pieces corresponds to multiplication. The modes of the operators $\mathcal{V}_{s_i}(z_i)$ correspond to vibrations of the cylinder pieces at different frequencies, and removing modes corresponds to reducing the number of frequencies at which the cylinder pieces can vibrate.}\label{string}
\end{figure}

In general, one can write the state of a one-dimensional chain of $N$ spins as
\begin{equation}
|\psi\rangle=\sum_{s_1,\ldots,s_N}\psi(s_1,\ldots,s_N)|s_1,\ldots,s_N\rangle,
\end{equation}
where $s_i$ is summed over all possible states of the $i$th spin and $\psi(s_1,\ldots,s_N)$ are complex coefficients. We shall here assume $s_i=\pm1$. For the family of states investigated in the present work, $\psi(s_1,\ldots,s_N)$ is chosen as
\begin{equation}\label{coef}
\psi(s_1,\ldots,s_N)=\langle\mathcal{V}_{s_1}(z_1)
\mathcal{V}_{s_2}(z_2)\ldots
\mathcal{V}_{s_N}(z_N)\rangle,
\end{equation}
where
\begin{multline}\label{vertex}
\mathcal{V}_{s_n}(z_n)\equiv
\exp\left(i\delta s_n\sqrt{\alpha}\phi_0
+s_n\sqrt{\alpha}\sum_{m=1}^M\frac{1}{\sqrt{m}}a_m^\dag z_n^m\right)\\
\times\exp\left(\delta s_n\sqrt{\alpha}\pi_0\ln(z_n)
-s_n\sqrt{\alpha}\sum_{m=1}^M\frac{1}{\sqrt{m}}a_m z_n^{-m}\right)
\end{multline}
and we shall always take $N$ to be even. Here, $\delta$ is either $0$ or $1$, $\alpha$ is a real and positive parameter, $z_n$ is a complex parameter, $\phi_0$, $\pi_0$, and $a_m$ are operators for which the nonvanishing commutation relations are $[\phi_0,\pi_0]=i$ and $[a_n,a_m^\dag]=\delta_{nm}$, the expectation value in \eqref{coef} is evaluated with respect to the state $|0\rangle$ defined by $\pi_0|0\rangle=a_m|0\rangle=0$, and we note that the resulting state is not necessarily normalized. Graphical illustrations of the MPS and iMPS constructions are given in Fig.~\ref{string}.

In the limit $M\rightarrow\infty$ and $\delta=1$, \eqref{vertex} is a chiral vertex operator of a conformal field theory with central charge $c=1$ \cite{cft} as investigated in \cite{cftimps}. In the following, we shall refer to the mode on which $\phi_0$ and $\pi_0$ act as the zero mode and use the term $m$th mode for the mode on which the bosonic annihilation operator $a_m$ acts. Note that $\delta=1$ ($\delta=0$) corresponds to keeping (omitting) the zero mode. We consider both cases because the zero mode has a special status and because we are particularly interested in the simplest situation $\delta=0$ and $M=1$, which involves only one mode.

Evaluating the expectation values, we find the explicit form
\begin{multline}\label{wave}
\psi(s_1,\ldots,s_N)=
\left[1-\delta+\delta\delta_s\left(\prod_{n<m}z_n^{\alpha s_ns_m}\right)\right]\\
\times\exp\left[-\alpha\sum_{n<m} s_ns_m\sum_{q=1}^M\frac{1}{q}\left(\frac{z_m}{z_n}\right)^q\right],
\end{multline}
where $\delta_s$ is defined to be unity if $\sum_{n=1}^Ns_n=0$ and zero otherwise. Note that the first factor in square brackets is the contribution from the zero mode, while the exponential is a product of $M$ factors coming from modes $1$ to $M$. In this and the next section, we choose $z_n=\exp(2\pi in/N)$, which gives a translationally invariant state for $\delta=0$. (Note that the definition in \eqref{coef} does not impose $|\psi\rangle$ to be translationally invariant in general.)

\subsection{Entropy}

To investigate the properties of the state \eqref{wave}, we first compute the Renyi entropy $S_L^{(2)}=-\log[\Tr(\rho_L^2)]$, where $\rho_L$ is the reduced density operator obtained by tracing out spins $L+1$ to $N$. A small calculation shows that
\begin{multline}\label{MCentropi}
e^{-S_L^{(2)}}=\\
\sum_{s_1,\ldots,s_N}\sum_{s'_1,\ldots,s'_N}
|\psi(s_1,\ldots,s_N)|^2|\psi(s'_1,\ldots,s'_N)|^2\\
\times\psi^*(s'_1,\ldots,s'_L,s_{L+1},\ldots,s_N)/\psi^*(s_1,\ldots,s_N)\\
\times\psi^*(s_1,\ldots,s_L,s'_{L+1},\ldots,s'_N)/\psi^*(s'_1,\ldots,s'_N)\\
\times\left[\sum_{s_1,\ldots,s_N}|\psi(s_1,\ldots,s_N)|^2\right]^{-2},
\end{multline}
where the stars mean complex conjugation. The right hand side of this equation has the form of an expectation value, and we compute it numerically by using the Metropolis Monte Carlo algorithm \cite{metropolis}.

\begin{figure}
\includegraphics[width=\columnwidth]{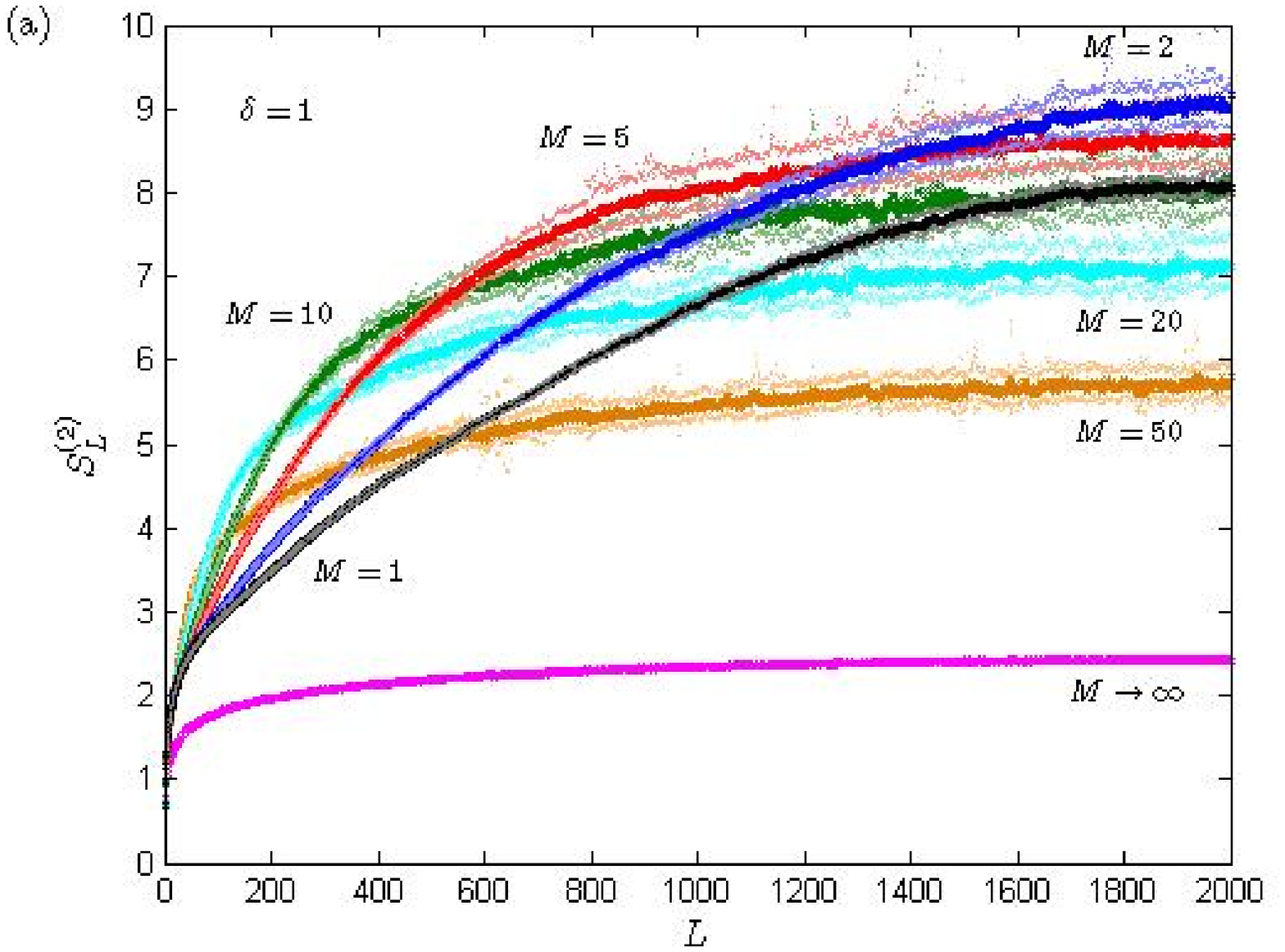}
\includegraphics[width=\columnwidth]{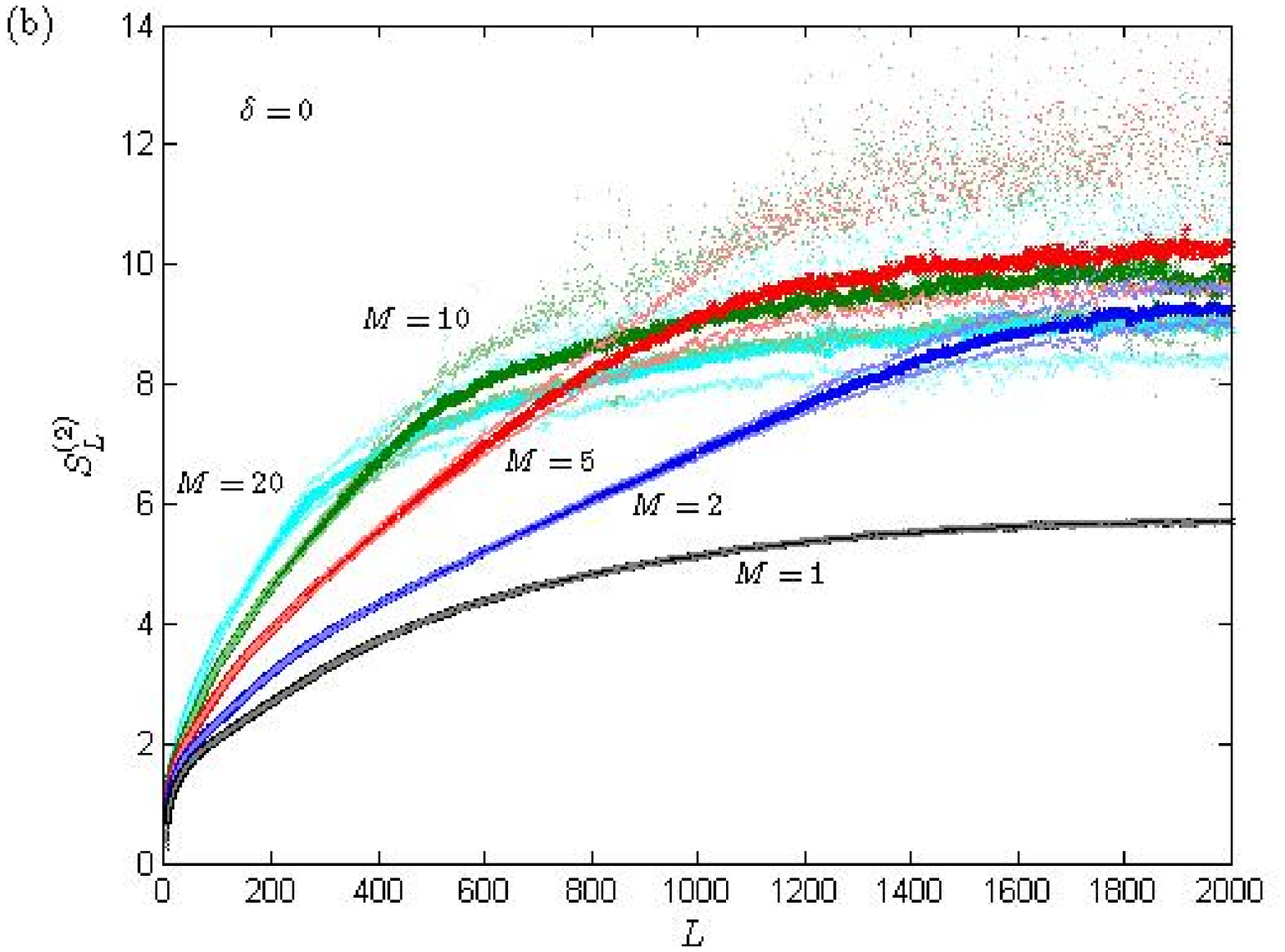}
\caption{(Color online) Renyi entropy $S_L^{(2)}$ of a subchain of length $L$ for $N=4000$, $\alpha=0.15$, various values of $M$, and $\delta=1$ (a) and $\delta=0$ (b) obtained from Monte Carlo simulations. The faint dots on both sides of the curves for $M=1,2,5,10,20,50$ indicate the uncertainty in the numerical estimates. These curves are computed by averaging the right hand side $r$ of \eqref{MCentropi} over the outcome of 50 or more Monte Carlo trajectories with different initial conditions, and the dots show $-\ln\left[\mean(r)\pm\sqrt{\var(r)}/2\right]$, where $\mean(r)$ and $\var(r)$ are the sample mean and the sample variance of $r$, respectively. We note that we use the same Monte Carlo trajectories to estimate $S_L$ for all values of $L$.}\label{entfig}
\end{figure}

For an MPS, the entropy of a subchain of length $L$ is limited by the logarithm of the virtual dimension, which is independent of $L$, and for critical systems the entropy grows approximately as $\ln(L)$ when the length of the subchain is short compared to the length of the complete chain \cite{cftimps}. The results for the state \eqref{wave} given in Fig.~\ref{entfig} shows that the entropy increases significantly, when the vertex operators are truncated to a finite number of modes. This is a bit surprising because we go from infinitely many modes representing infinite dimensional Hilbert spaces to a finite number of modes representing infinite dimensional Hilbert spaces and shows that a single bosonic mode is, in fact, sufficient to obtain states, which contain a significant amount of entropy. We also observe that qualitatively similar results are obtained if the zero mode is or is not included.

\begin{figure}
\includegraphics[width=\columnwidth]{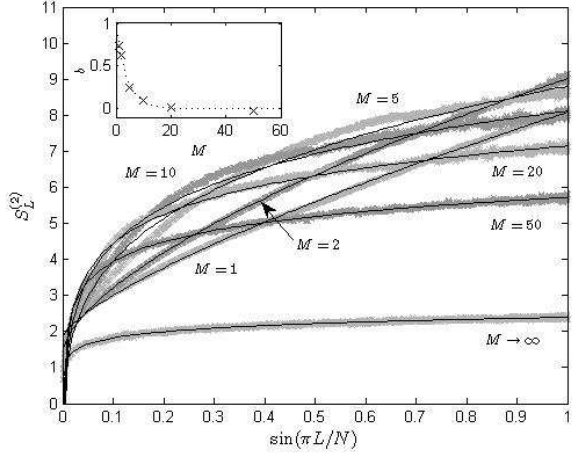}
\caption{The same data as in Fig.~\ref{entfig}(a), but plotted as a function of $\sin(\pi L/N)$. The black solid lines are fits of the form $S_L^{(2)}=a[\sin(\pi L/N)]^b+c$ to the numerical data (gray crosses), where $a$, $b$, and $c$ are fitting parameters. The inset shows the exponent $b$ (crosses) as a function of $M$ ($b=0.0093$ for $M\rightarrow\infty$), and the dotted line is the fit $b=0.967\exp(-0.246M)$.}\label{entMfits}
\end{figure}

\begin{figure}
\includegraphics[width=0.6\columnwidth]{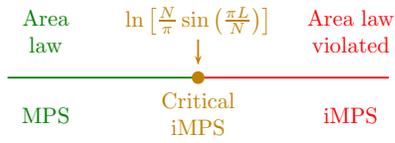}
\caption{(Color online) Schematic view of the scaling of the entropy of a subchain with the length of the subchain for various models. For MPS the entropy is bounded by the logarithm of the virtual dimension, which is independent of $L$. For the critical case, the entropy scales with $\ln[N\sin(\pi L/N)/\pi]$, which is on the boundary between constant behavior and a power law, and for the iMPS considered in the present paper, there is a transition from the critical case to a power law scaling of the entropy with $\sin(\pi L/N)$ as $M$ decreases towards one.}\label{fcimps}
\end{figure}

In addition to the pictorial representation of the transition from $M=1$ to $M\rightarrow\infty$ in Fig.~\ref{entfig}(a), it is interesting to investigate the approximate scaling behavior of $S_L^{(2)}$ with $L$ as a function of $M$. Inspired by the critical case $M\rightarrow\infty$, for which the entropy is expected to follow the relation $S_L^{(2)}=\ln(N\sin(\pi L/N)/\pi)/4+\textrm{constant}$ \cite{cftimps}, we plot the entropy as a function of $\sin(\pi L/N)$ in Fig.~\ref{entMfits}. For each value of $M$ we fit a power law of the form $S_L^{(2)}=a(\sin(\pi L/N))^b+c$ to the data, where $a$, $b$, and $c$ are fitting parameters, and we plot the exponent $b$ as a function of $M$. The fits show that the power law is a good approximation when $M$ is small provided the subchain does not consist of only a few spins. For intermediate $M$'s there are some discrepancies, and for $M=50$ and $M\rightarrow\infty$, the exponent is close to zero. This reflects the transition to a logarithmic scaling for $M\rightarrow\infty$. Specifically, we find that $S_L^{(2)}=0.246\ln(\sin(\pi L/N))+2.39$ provides a good fit to the data for $M\rightarrow\infty$, which is in accordance with the expression given above. Figure~\ref{fcimps} provides a schematic summary of the scaling behavior of the entropy for MPS, the critical iMPS studied in \cite{cftimps}, and the iMPS considered in the present article.

\begin{figure}
\includegraphics[width=\columnwidth]{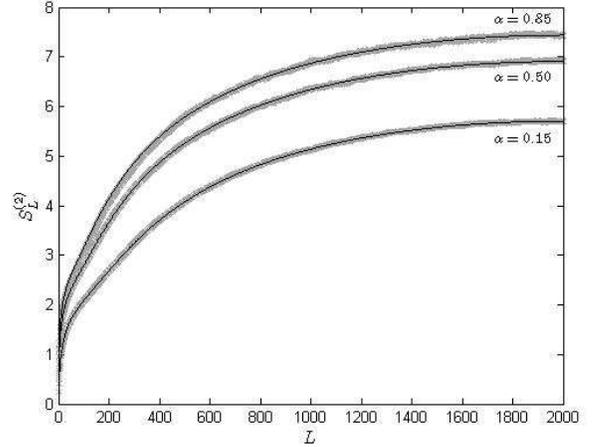}
\caption{Renyi entropy $S_L^{(2)}$ of a subchain of length $L$ for $N=4000$, $M=1$, $\delta=0$, and various values of $\alpha$ as indicated. The gray crosses are obtained from Monte Carlo simulations, and the black solid lines show the values predicted by the approximate relation in Eq.~\eqref{entropy}.}\label{alphaunent}
\end{figure}

The dependence of the entropy on $\alpha$ is exemplified in Fig.~\ref{alphaunent} for $M=1$ and $\delta=0$, where the entropy is observed to increase with $\alpha$, and a very similar behavior is found for $M=1$ and $\delta=1$. We can hence use $\alpha$ to fine-tune the entropy to a desired value. In Appendices \ref{app1} and \ref{app2}, we show that the entropy for $M=1$ and $\delta=0$ is given approximately by the relation
\begin{equation}\label{entropy}
S_L^{(2)}=-\ln\left(\sqrt{\frac{\det(M_1+M_3)}{\det(M_1+M_2+M_3)}}\right)
\end{equation}
in the limit of large $N$, where $M_1$, $M_2$, and $M_3$ are the eight-by-eight matrices defined in Eqs.~\eqref{M1}, \eqref{M2}, and \eqref{M3}. To obtain this result, we use that $z_n=\exp(2\pi in/N)$ has practically the same value for several nearby values of $n$ when $N$ is large and that the sum of $s_n$ for such a group of $n$-values approximately follow a Gaussian probability distribution when spin configurations are chosen at random. The expression \eqref{entropy} is also plotted in the figure, and we observe an excellent agreement with the numerical results.

\subsection{Correlation function}

Another important characteristic is the behavior of the correlation function between two operators acting on different spins. Specifically, we shall consider
\begin{equation}\label{correlator}
\langle\sigma_z(0)\sigma_z(k)\rangle=
\frac{\sum_{s_1,\ldots,s_N}s_0s_k|\psi(s_1,\ldots,s_N)|^2} {\sum_{s_1,\ldots,s_N}|\psi(s_1,\ldots,s_N)|^2},
\end{equation}
where $\sigma_z(k)$ is the third of the Pauli operators acting on the $k$th spin, $\sigma_z(0)=\sigma_z(N)$, and the expectation value is computed with respect to the atomic state \eqref{wave}. Note that $|\psi(s_1,\ldots,s_N)|^2$ is translationally invariant, and hence $\langle\sigma_z(n)\sigma_z(n+k)\rangle=\langle\sigma_z(0)\sigma_z(k)\rangle$, where $n$ is an integer, and the argument of $\sigma_z$ is understood to be modulus $N$.

\begin{figure}
\includegraphics[width=\columnwidth]{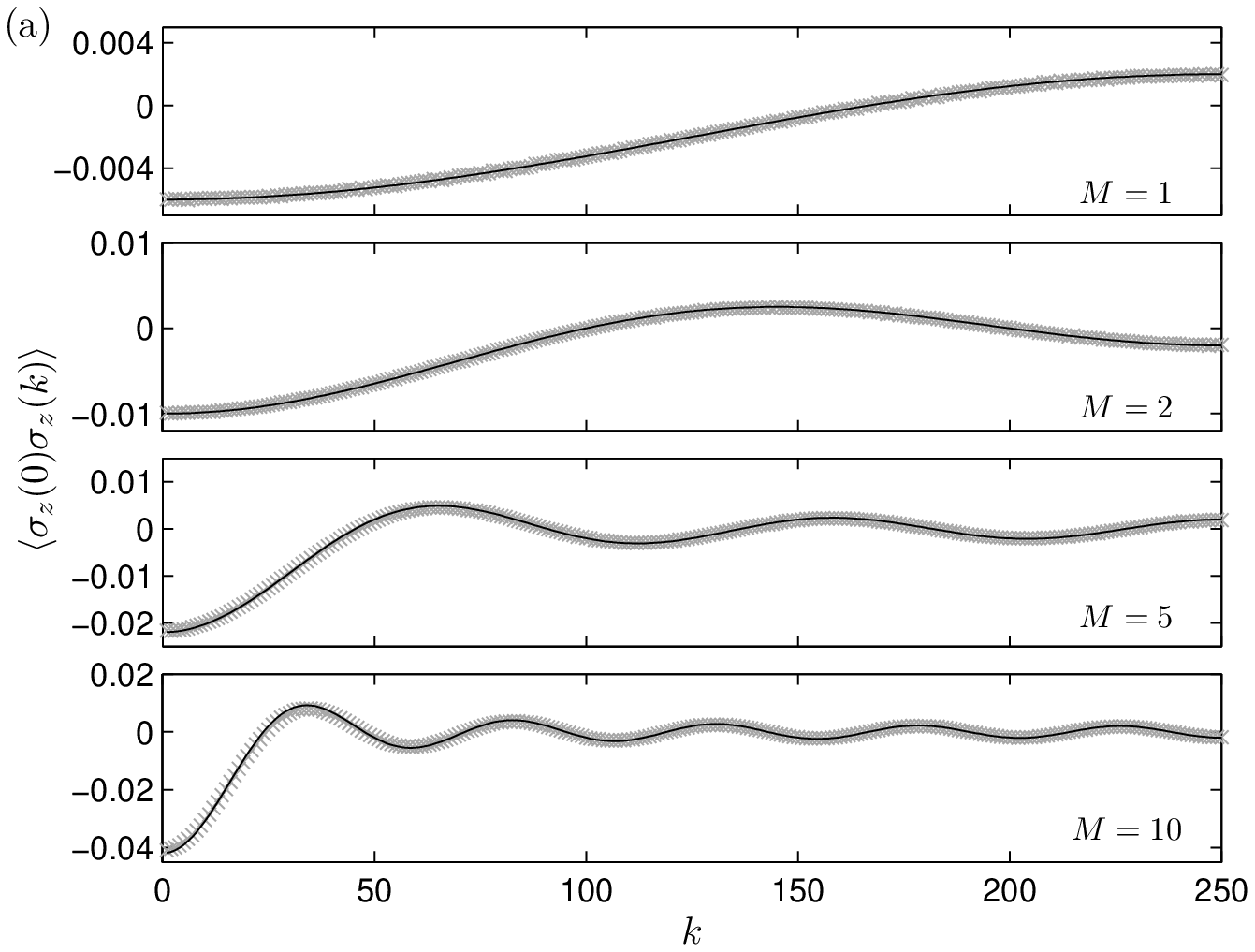}
\includegraphics[width=\columnwidth]{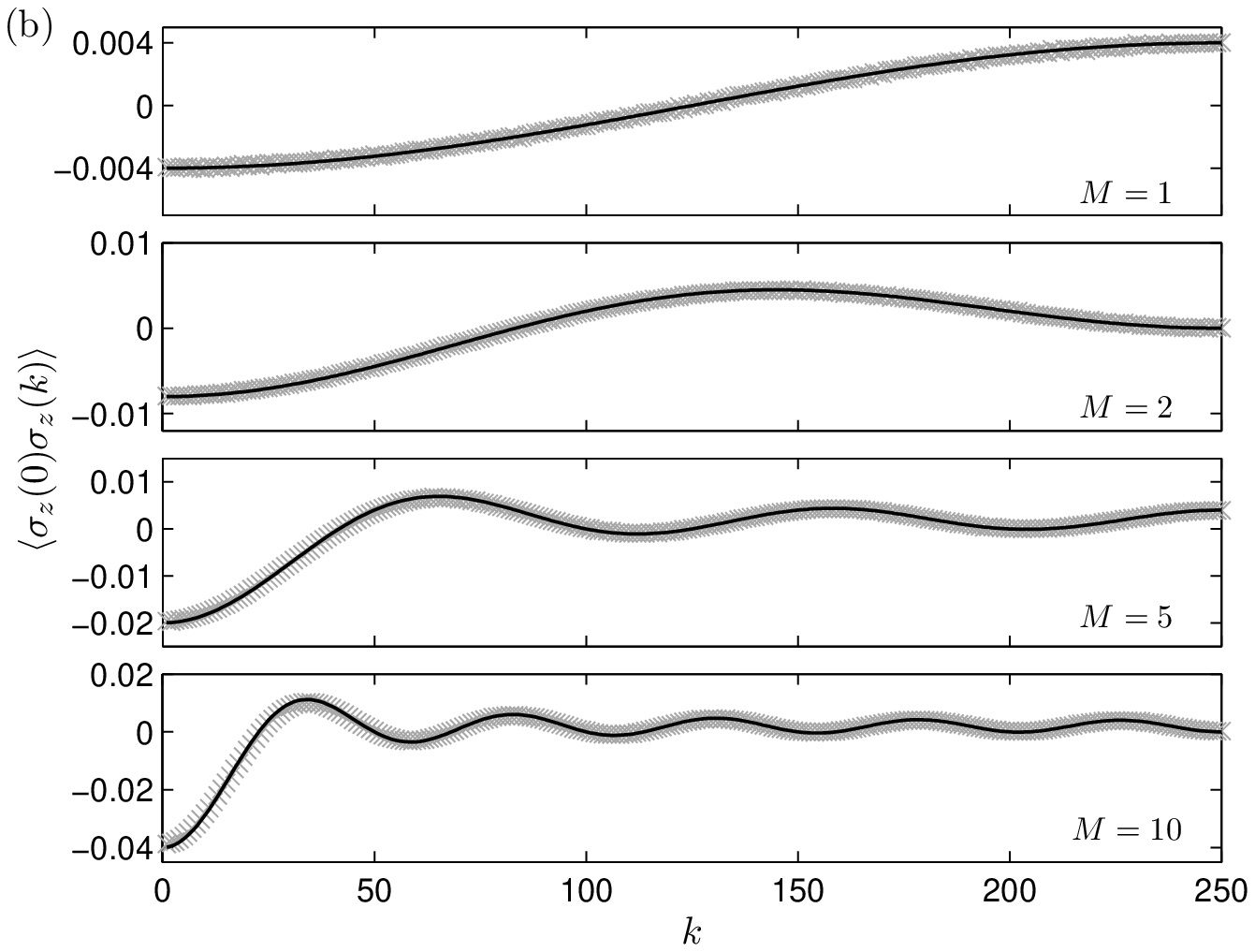}
\caption{The correlation function \eqref{correlator} as a function of the separation $k$ between the two spins for $N=500$, $\alpha=0.15$, and $\delta=1$ (a) and $\delta=0$ (b). The gray crosses show results obtained from Monte Carlo simulations, and the solid black lines are the approximation in Eq.~\eqref{corM}. (Note that the point $k=0$, for which the correlation function is unity, is not included in the plots.)}\label{corfig}
\end{figure}

Numerical results for the correlation function are given in Fig.~\ref{corfig} for various values of $M$, and we observe an excellent agreement with the approximate relation (for $k\neq N$ modulus $N$) obtained in App.~\ref{app4}
\begin{equation}\label{corM}
\langle\sigma_z(0)\sigma_z(k)\rangle=
-\frac{\delta+2\sum_{m=1}^M\cos(2m\pi k/N)}{N},
\end{equation}
which is expected to be valid for $M\ll N$. It is interesting to note that the correlation function for $M=1$ does not decay with the separation $k$ for fixed $N$, but instead exhibits an oscillatory behavior with a wavelength, which is equal to the length of the chain. This is a very different behavior from the critical case $M\rightarrow\infty$ and $\delta=1$, in which the correlation function shows an antiferromagnetic ordering and a magnitude decreasing with $k$ \cite{cftimps}. It is also in contrast to MPS, for which the correlation function decays exponentially with the separation \cite{mpsreview}. We also note that the correlation functions for $\delta=1$ and $\delta=0$ only differ by the constant term $-\delta/N$, and both correlation functions do not depend on $\alpha$.

\section{Experimental implementation}\label{implementations}

\subsection{Method}

The considered class of states is also interesting from the point of view that the structure of Eq.\ \eqref{coef} and the fact that $\mathcal{V}_{s_n}(z_n)$ only depends on the state of the $n$th spin provide a direct recipe for preparing the states experimentally in which an ancilla system sequentially interacts once with each of the spins. The form of the truncated vertex operators furthermore allows us to decompose the ancilla system into $M+\delta$ independent bosonic modes that can be treated separately. Specifically, \eqref{coef} and \eqref{vertex} imply
\begin{multline}\label{opeprod}
|\psi\rangle=\left\langle\prod_{n=1}^Ne^{i\delta\sigma_z(n)\sqrt{\alpha}\phi_0}
e^{\delta\sigma_z(n)\sqrt{\alpha}\pi_0\ln(z_n)}\right\rangle\\
\times\prod_{m=1}^M\Bigg\langle\prod_{n=1}^N
e^{\sigma_z(n)\sqrt{\alpha}m^{-1/2}a_m^\dag z_n^m}
e^{-\sigma_z(n)\sqrt{\alpha}m^{-1/2}a_m z_n^{-m}}\Bigg\rangle\\
\times\sum_{s_1,\ldots,s_N}|s_1,\ldots,s_N\rangle,
\end{multline}
where $\prod_{n=1}^No_n$ for some set of operators $o_n$ is understood to mean $o_1o_2\ldots o_N$, and it is hence possible to prepare the desired state from a simple initial state if one can implement the $M+\delta$ operators in \eqref{opeprod}.

Let us first consider the operator coming from the $m$th mode. This operator is proportional to
\begin{equation}
\Bigg\langle\prod_{n=1}^N
\exp\left[\sigma_z(n)\sqrt{\alpha}m^{-1/2}\left(a_m^\dag z_n^m
-a_m z_n^{-m}\right)\right]\Bigg\rangle,
\end{equation}
which for $z_n=\exp(2\pi in/N)$ is the expectation value of a product of $N$ displacement operators, where the displacement $\sigma_z(n)\sqrt{\alpha}m^{-1/2}\exp(2\pi inm/N)$ accomplished by the $n$th displacement operator depends on the state of the $n$th spin. In other words, we can implement the contribution from the $m$th mode by initializing the bosonic mode in the vacuum state, engineering the Hamiltonian
\begin{equation}
H=i\hbar\sigma_z(n)(ga_m^\dag-g^*a_m),
\end{equation}
where $g$ is a coupling strength, applying it for a time $t$ such that $\int_0^t gdt=\sqrt{\alpha}m^{-1/2}\exp(2\pi inm/N)$ for each of the spins, and finally projecting the bosonic mode on the vacuum state via a conditional measurement.

\begin{figure}
\includegraphics[width=\columnwidth]{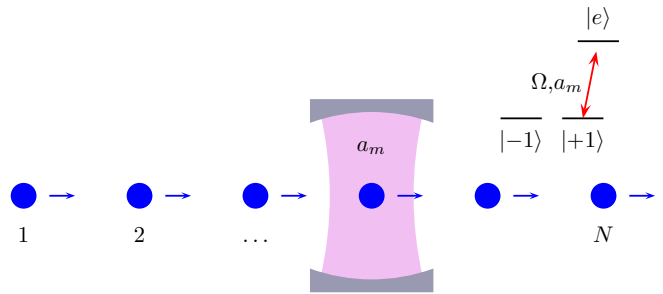}
\caption{(Color online) To prepare the state \eqref{wave} experimentally, we initialize $N$ spins, which are here encoded in atoms, in the state $(|{+1}\rangle+|{-1}\rangle)^{\otimes N}$, and for each mode included in the vertex operators we let the spins interact one by one with a cavity field, which is subsequently subjected to a measurement. The inset shows the level structure of the atoms and the interaction with the cavity field and a classical field as described further in the text.}\label{cavity}
\end{figure}

The bosonic mode could be a single mode of a cavity, and the spins could be atoms traversing the cavity one by one as illustrated in Fig.~\ref{cavity}. The movements of the atoms in and out of the cavity could, for instance, be accomplished by the method demonstrated in \cite{conveyor}, where laser fields are used to form a conveyor belt for the atoms. The two levels of the spins correspond to two ground state levels $|{\pm1}\rangle$ of the atoms, and we assume that the cavity field couples only the state $|{+1}\rangle$ to an excited state. By choosing the cavity field to be far detuned from the atomic transition and driving the same transition with a classical field at the same frequency, one achieves the Hamiltonian
\begin{equation}
H=-\frac{\hbar}{\Delta}(|\Omega|^2+\tilde{g}\Omega a_m^\dag+\tilde{g}^*\Omega^*a_m+|\tilde{g}|^2a_m^\dag a_m)|{+1}\rangle\langle{+1}|,
\end{equation}
where $\Delta$ is the detuning between the atomic transition and the frequency of the cavity mode, $\tilde{g}$ is the coupling strength between the cavity mode and the atom, and $\Omega$ is the Rabi frequency for the driving of the atom with the classical field. The ac stark shift $-\hbar|\Omega|^2|{+1}\rangle\langle{+1}|/\Delta$ can be compensated by coupling $|{+1}\rangle$ to another excited state with another classical field, and the term $-\hbar|\tilde{g}|^2a_m^\dag a_m|{+1}\rangle\langle{+1}|/\Delta$ can be neglected if $\Omega$ is sufficiently large. The resulting interaction only displaces the amplitude of the cavity field if the atom is in the state $|{+1}\rangle$. One way to correct for this is to apply a displacement operation to the cavity field, which is independent of the state of the atom. This is, in fact, not necessary, however, because the error in the displacement occurring when the $n$th atom interacts with the cavity field is opposite to the error in the displacement occurring when the $(n+N/2)$th atom interacts with the cavity field. Except for an overall phase factor, the only consequence of not applying the additional displacement operations is to introduce some single atom phase factors, but these can be canceled by applying the unitary operator
\begin{equation}
\exp\left\{2i\alpha\sum_{n=1}^N[\sigma_z(n)+1]\Imag\left(\sum_{m=1}^{n-1}e^{2\pi i(n-m)/N}\right)\right\}
\end{equation}
to the atoms after all the interactions with the cavity field have taken place. The final projection of the cavity field on the vacuum state may, for instance, be accomplished with a photo detector, or one can use atoms to probe the state of the cavity field (see \cite{haroche} for an implementation in the microwave regime).

The contribution from the zero mode (if it is included) can be implemented along similar lines. The key point is to note that
\begin{multline}
\delta_s\prod_{n<m}z_n^{\alpha s_ns_m}=\lim_{\beta\rightarrow\infty}
\Bigg\langle\prod_{n=1}^N\exp\Big[\sqrt{\beta}s_n\\
\times\left(a_0^\dag e^{\alpha\pi in/(\beta N)}
-a_0 e^{-\alpha\pi in/(\beta N)}\right)\Big]\Bigg\rangle,
\end{multline}
where $a_0$ is a bosonic annihilation operator and the expectation value is evaluated in the vacuum state of that operator. We can hence use the same procedure as before if we take the limit where $|{\int\tilde{g}\Omega dt}|$ goes to infinity while the deviation of the phase of $\int\tilde{g}\Omega dt$ from $-\pi/2$ goes to zero as $1/|{\int\tilde{g}\Omega dt}|^2$. The physical reason why a $\delta_s$ appears in this limit is that the norm of the sum of the displacements is either zero or infinite when the amplitude of the classical field is infinite, and there is only a nonvanishing overlap with the vacuum state of the cavity mode after the interactions in the former case.

We finally need a way to initialize the atoms in the state $\sum_{s_1,\ldots,s_N}|s_1,\ldots,s_N\rangle$. This can be done by pumping all the atoms to the state $|{-1}\rangle$ and applying a $\pi/2$ pulse between the levels $|{-1}\rangle$ and $|{+1}\rangle$.

\subsection{Success probability}

\begin{figure}
\includegraphics[width=\columnwidth]{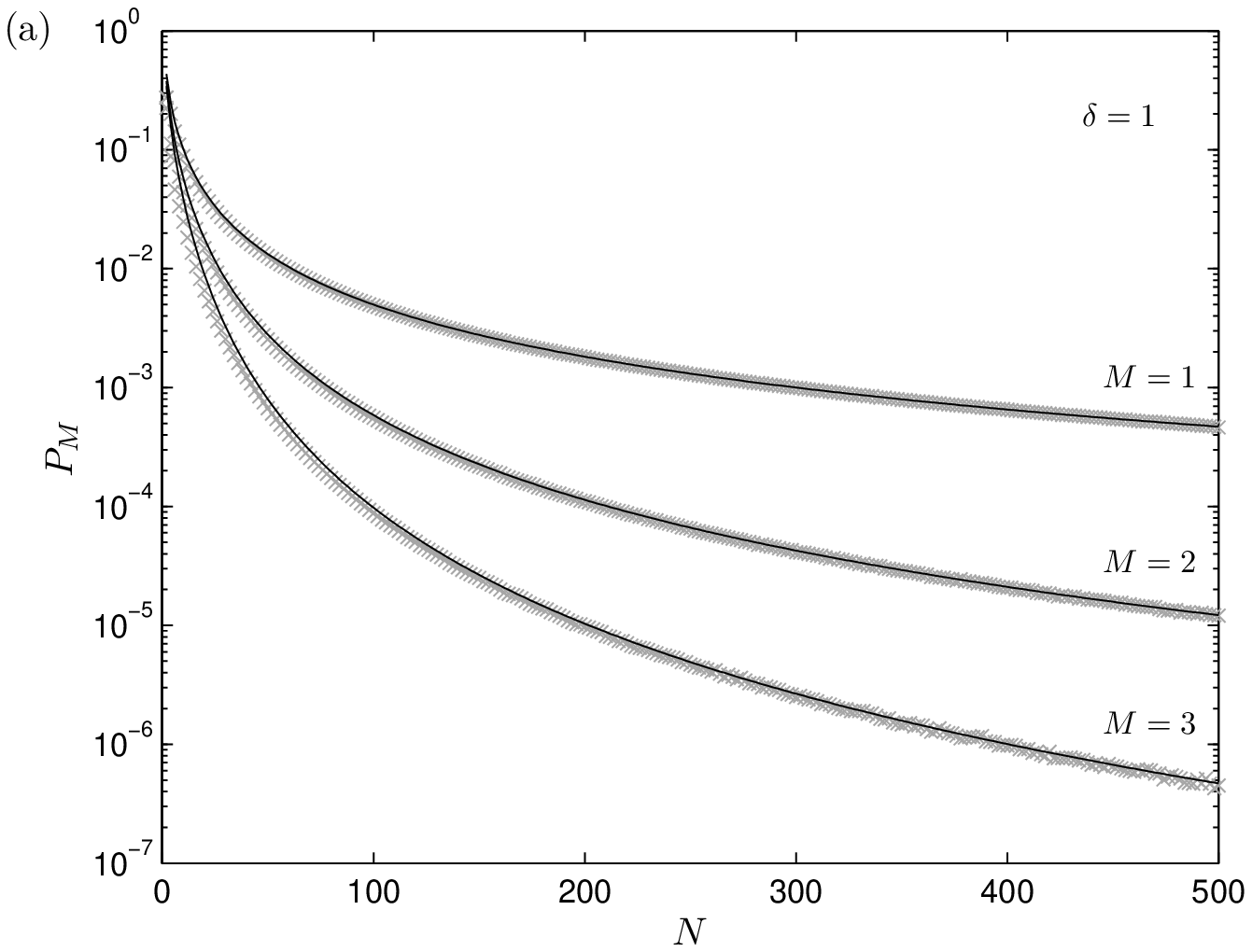}
\includegraphics[width=\columnwidth]{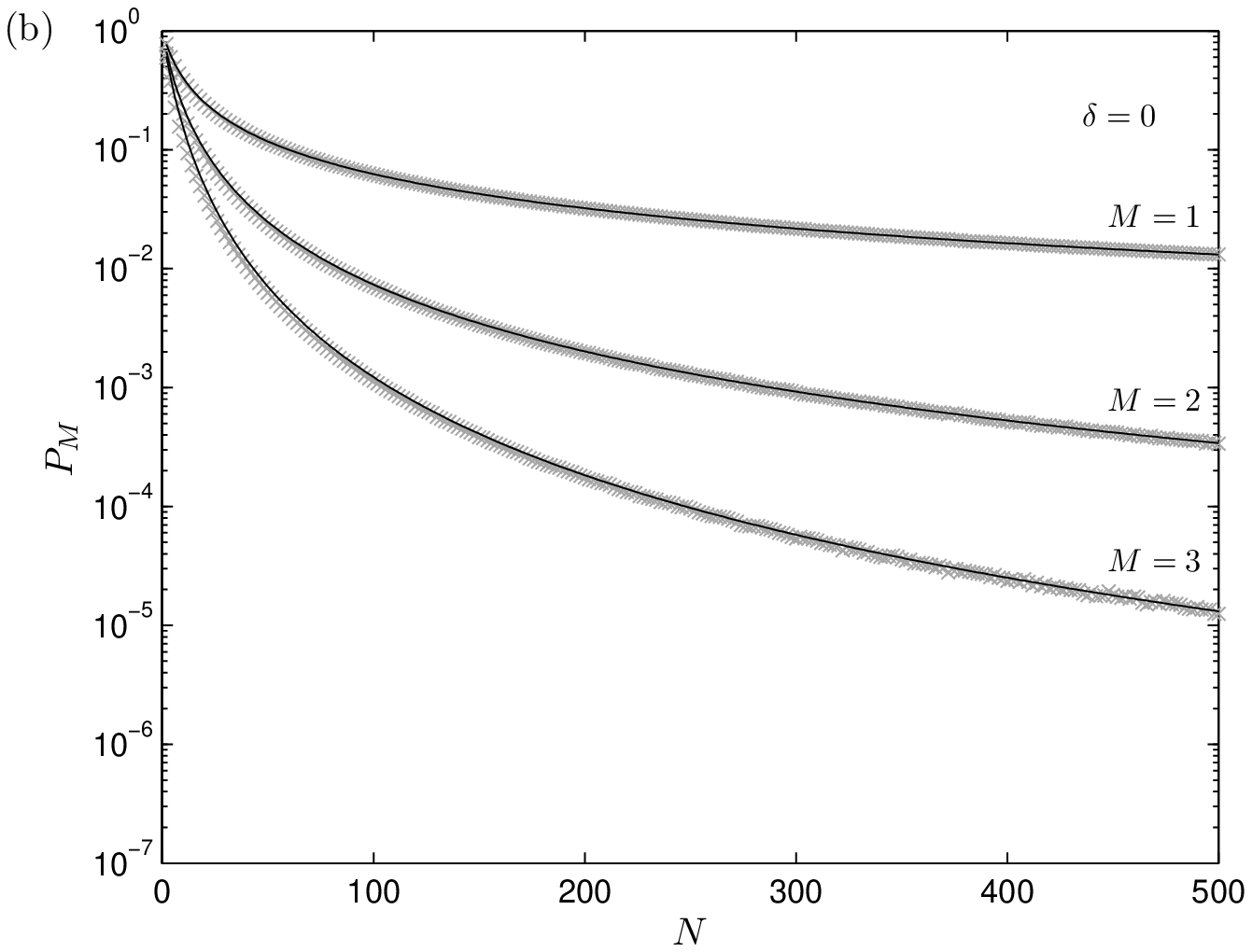}
\caption{Probability of successfully projecting all the $M+\delta$ cavity modes on the vacuum state after the interactions with the atoms as a function of the number of atoms for $\alpha=0.15$ and $\delta=1$ (a) and $\delta=0$ (b). The gray crosses are numerical results, and the black solid lines represent the approximate relation in Eq.~\eqref{PM}.}\label{Mprob}
\end{figure}

The proposed generation scheme is only efficient if the probability $P_M$ for successfully projecting all the cavity modes on the vacuum state after the interactions with the atoms is not too small. $P_M$ is the square of the norm of the final state of the atoms relative to the square of the norm of the initial state of the atoms, and we find
\begin{equation}\label{P}
P_M=\frac{1}{2^N}\sum_{s_1,\ldots,s_N}(1-\delta+\delta\delta_s)
\exp\left(-\alpha N^2\sum_{q=1}^M\frac{1}{q}\left|\tilde{s}_q\right|^2\right),
\end{equation}
where
\begin{equation}\label{tildesq}
\tilde{s}_q\equiv\frac{1}{N}\sum_{n=1}^N s_ne^{2\pi iqn/N}.
\end{equation}
We evaluate the expression in \eqref{P} numerically by including a larger and larger number of randomly chosen configurations of the spins in the sum until the sum divided by the number of included configurations converges to a constant value, and the results are shown in Fig.~\ref{Mprob}.

It is possible to obtain an analytical approximation to $P_M$ as follows. In App.~\ref{app1} we show that the values of $x_N=N\Real(\tilde{s}_1)$ and $y_N=N\Imag(\tilde{s}_1)$ for randomly chosen spin configurations in the limit of large $N$ follow a Gaussian probability distribution
\begin{equation}
P(x_N,y_N)=\frac{1}{\pi N}\exp\left[-(x_N^2+y_N^2)/N\right],
\end{equation}
and the same distribution also applies to the real and imaginary parts of $N\tilde{s}_q$. Hence, if we start from the atomic state $\sum_{s_1,\ldots,s_N}|s_1,\ldots,s_N\rangle$ and only apply the operations required to take the $q$th mode in the truncated vertex operators into account, the projection on the vacuum state for that mode proceeds successfully with probability
\begin{multline}
\int_{-\infty}^\infty dx^{(q)}_N\int_{-\infty}^\infty dy^{(q)}_N
\exp\left\{-\frac{\alpha}{q}\left[\left(x_N^{(q)}\right)^2
+\left(y_N^{(q)}\right)^2\right]\right\}\\
\times P(x_N^{(q)},y_N^{(q)})=\left(1+\frac{\alpha N}{q}\right)^{-1}.
\end{multline}
Since the inclusion of the zero mode projects the atomic state on the subspace of states with $\sum_{n=1}^Ns_n=0$ in addition to introducing some phase factors, the probability of success when we only include the zero mode is $2^{-N}\times N!/[(N/2)!]^2$, which by use of Stirling's approximation can be replaced by $[2/(\pi N)]^{1/2}$ for large $N$.

From the arguments in App.~\ref{app3} we expect the distributions of $N\tilde{s}_q$ for a moderate number of different values of $q$ to be approximately independent in the limit of large $N$. This means that the success probability $P_M$ factorizes into a product of the individual contributions from the $M+\delta$ modes, and we hence predict the approximate relation
\begin{equation}\label{PM}
P_M=\left(1-\delta+\delta\sqrt{\frac{2}{\pi N}}\right)
\prod_{q=1}^M\left(1+\frac{\alpha N}{q}\right)^{-1}.
\end{equation}
This result is also plotted in Fig.~\ref{Mprob}, and we see an excellent agreement with the numerical computations even for only moderately high $N$. Equation \eqref{PM} suggests that $P_M$ scales approximately as $N^{-(M+\delta/2)}$, and for moderate $M$ we hence conclude that the number of operations required to prepare the desired state grows only polynomially in $N$. The proposed method can, however, not be used for generating the critical state investigated in \cite{cftimps} since the success probability vanishes in the limit $M\rightarrow\infty$.

\section{Modifications}\label{modifications}

The experimental preparation scheme outlined above can be modified in various ways, which lead to new states that extend the family of states considered above, and we investigate the effects of two such modifications in the following. We focus on the case of a single bosonic mode, i.e., $M=1$ and $\delta=0$, throughout.

\subsection{Squeezing of the bosonic mode before and after the interactions}\label{squeezing}

One way to modify the preparation scheme is to modify the state of the cavity field before and after the interactions with the atoms. A nontrivial example is to apply a squeezing operation, i.e., the time evolution operator $\exp(r(e^{2i\phi}(a_1^\dag)^2-e^{-2i\phi}a_1^2)/2)$, where $r$ and $\phi$ are real parameters, before the interactions and to apply the inverse operation after the interactions. Physically, this can be done by placing a nonlinear crystal inside the cavity and pumping it with a classical field \cite{loudon}, and mathematically this amounts to replacing the vacuum expectation value in Eq.~\eqref{coef} by an expectation value with respect to the squeezed vacuum state
\begin{equation}
|\textrm{sq}\rangle=(1-4|A|^2)^{1/4}e^{A(a_1^\dag)^2}|0\rangle,
\end{equation}
where $A=e^{2i\phi}\tanh(r)/2$, i.e.,
\begin{multline}
\psi(s_1,\ldots,s_N)=\langle\textrm{sq}|\prod_{n=1}^N
e^{s_n\sqrt{\alpha}a_1^\dag\exp(2\pi in/N)}\\
\times e^{-s_n\sqrt{\alpha}a_1\exp(-2\pi in/N)}|\textrm{sq}\rangle\\
=\exp\Bigg[-\alpha\sum_{n<m}s_ns_me^{2\pi i(m-n)/N}\\
-\frac{4\alpha N^2|A|^2}{1-4|A|^2}|\tilde{s}_1|^2
+\frac{2\alpha N^2}{1-4|A|^2}\Real(A^*\tilde{s}_1^2)\Bigg],
\end{multline}
where $\tilde{s}_1$ is defined in Eq.~\eqref{tildesq} and we have used a theorem derived in \cite{sqmath} to evaluate the expectation value.

\begin{figure}
\includegraphics[width=\columnwidth]{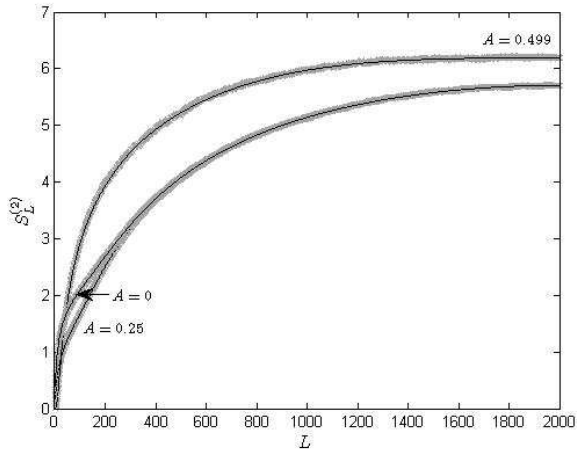}
\caption{Renyi entropy $S_L^{(2)}$ as a function of the length of the subchain for $N=4000$, $\alpha=0.15$, and various values of squeezing of the cavity field. The gray crosses are results from Monte Carlo simulations, and the black solid lines are the analytical approximation derived in App.~\ref{app2}.}\label{sqent}
\end{figure}

\begin{figure}
\includegraphics[width=\columnwidth]{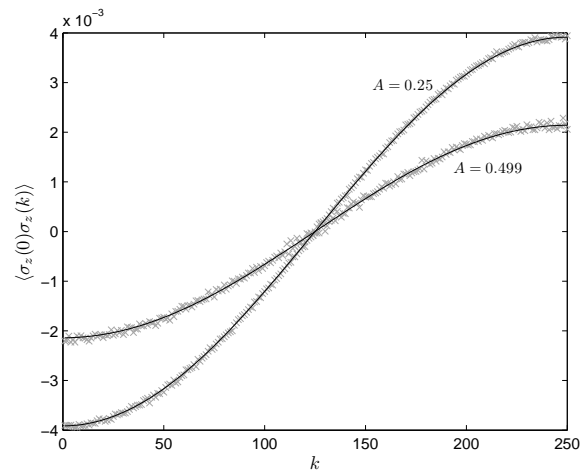}
\caption{The correlation function \eqref{correlator} as a function of the separation between the spins for $N=500$, $\alpha=0.15$, and various values of the squeezing of the cavity field. The gray crosses are results from Monte Carlo simulations, and the black solid lines are the analytical approximation in Eq.~\eqref{sqacor}.}\label{sqcor}
\end{figure}

We can use the same approach as for $A=0$ to derive analytical approximations to the Renyi entropy (see App.~\ref{app2}), the correlation function (see App.~\ref{app4})
\begin{multline}\label{sqacor}
\langle\sigma_z(0)\sigma_z(k)\rangle=-\cos(2\pi k/N)\\
\times\frac{2\alpha (1+4|A|^2)/(1-4|A|^2)+2\alpha^2N}
{1+2\alpha N(1+4|A|^2)/(1-4|A|^2)+\alpha^2N^2}
\end{multline}
(for $k\neq N$ modulus $N$), and the success probability
\begin{multline}\label{psq}
P_\textrm{sq}=\frac{1}{2^N}\sum_{s_1,\ldots,s_N}\exp\bigg[
-\alpha N^2\frac{1+4|A|^2}{1-4|A|^2}|\tilde{s}_1|^2\\
+\alpha N^2\frac{4}{1-4|A|^2}\Real\left(A^*\tilde{s}_1^2\right)\bigg]\\
=\left(1+2\alpha N\frac{1+4|A|^2}{1-4|A|^2}+\alpha^2N^2\right)^{-1/2}.
\end{multline}
Numerical and analytical results for the entropy and the correlation function are provided in Figs.~\ref{sqent} and \ref{sqcor}, and it is apparent that $|A|$ needs to be close to $1/2$ in order to observe significant deviations from the case $A=0$. More precisely, Eqs.~\eqref{sqacor} and \eqref{psq} show that $(1+4|A|^2)/(1-4|A|^2)$ should be on the order of $\alpha N$ or larger for the squeezing to have a significant effect. If $|A|$ is close to $1/2$, we observe an increase in the entropy compared to the case $A=0$. The correlation function keeps the same shape but the amplitude decreases from $2/N$ to $1/N$ as $A$ increases from $0$ to $1/2$. Finally, we note that $P_\textrm{sq}\sim \sqrt{\epsilon/(\alpha N)}$ when $|A|=1/2-\epsilon$ and $\epsilon\ll1/(\alpha N)$.

\subsection{Projection of a part of the bosonic mode on vacuum between each interaction}\label{projection}

Another natural way to modify the preparation scheme is to allow a bit of the cavity field to leak out between each interaction with one atom and project the field that has escaped from the cavity on the vacuum state to remain within the set of pure states. The leakage corresponds to the application of a beam splitter operation, and hence
\begin{multline}
\psi_\textrm{pr}(s_1,\ldots,s_N)=\Bigg\langle\prod_{n=1}^N
e^{g^*a_1b_n^\dag-ga_1^\dag b_n}\\
\times e^{\sqrt{\alpha}s_na_1^\dag\exp(2\pi in/N)}
e^{-\sqrt{\alpha}s_na_1\exp(-2\pi in/N)}\Bigg\rangle,
\end{multline}
where $b_n$ are bosonic annihilation operators, i.e., $[b_n,b_m^\dag]=\delta_{nm}$, and $g$ is a parameter quantifying the amount of leakage (the relative loss in intensity is $\sin^2(|g|)$). Commuting operators in this expression, one can show that it is again possible to only displace the cavity field each time an atom is in the state $|{+1}\rangle$ provided (i) one applies appropriate displacement operations to the field that leaks out of the cavity, (ii) one applies an appropriate displacement to the cavity field after completing the interaction with all the atoms and before the projection on the vacuum state, and (iii) one applies appropriate single atom phase shifts to the atoms. We note that the displacements of the field amplitudes can be accomplished by use of a strong classical field and a beam splitter as proposed in \cite{displace}.

\begin{figure}
\includegraphics[width=\columnwidth]{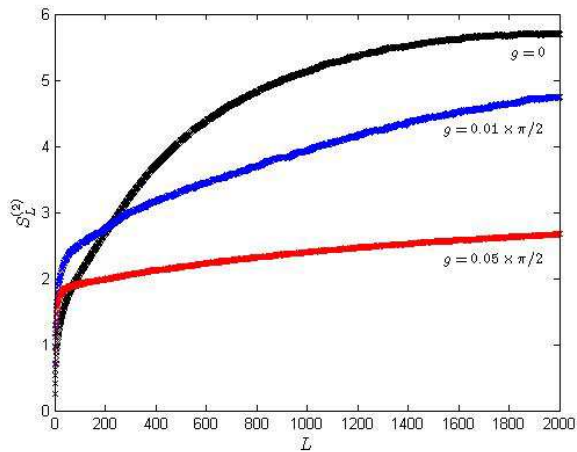}
\caption{(Color online) Renyi entropy $S_L^{(2)}$ obtained from Monte Carlo simulations with $N=4000$, $\alpha=0.15$, and various amounts of leakage.}\label{dkent}
\end{figure}

\begin{figure}
\includegraphics[width=\columnwidth]{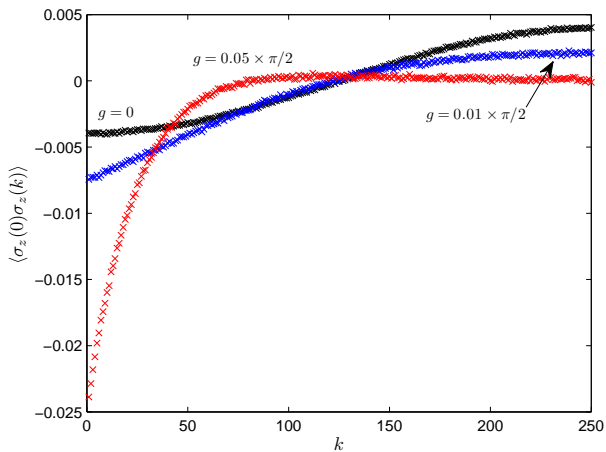}
\caption{(Color online) The correlation function $\langle\sigma_z(0)\sigma_z(k)\rangle$ obtained from Monte Carlo simulations with $N=4000$, $\alpha=0.15$, and various amounts of leakage.}\label{dkcor}
\end{figure}

Evaluating the expectation value, we find that the wave function
\begin{multline}
\psi_\textrm{pr}(s_1,\ldots,s_N)=\\
\prod_{n<m}\exp\left\{-\alpha s_ns_m[\cos(|g|)]^{m-n}e^{2\pi i(m-n)/N}\right\}
\end{multline}
is, in fact, again given by Eq.~\eqref{wave} (with $M=1$ and $\delta=0$), but now $z_n=\cos(|g|)^ne^{2\pi in/N}$. The numerical results for the entropy and the correlation function \eqref{correlator} in Figs.~\ref{dkent} and \ref{dkcor} show that even a small leakage has a significant impact on the properties of the produced states. The entropy is reduced significantly, and the correlation function now decays with separation. For $g=0.05\times\pi/2$, the decay is roughly exponential. This indicates that it is important to exclude losses from the setup to obtain the high entropies and long range correlations found in Sec.~\ref{Mmodes}. The correlation function also shows that antialignment of nearby spins become more favorable when leakage is included. This can be understood from the fact that the field leaking out of the cavity is more likely to be projected on the vacuum state if the total displacement of the cavity field is close to zero at all times during the preparation.

\begin{figure}
\includegraphics[width=\columnwidth]{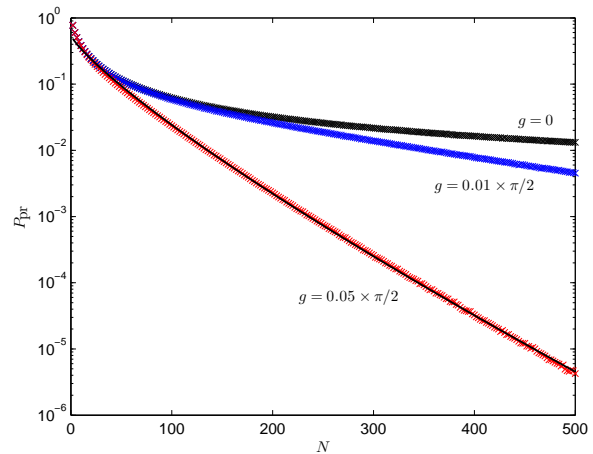}
\caption{(Color online) Success probability as a function of the number of spins for $\alpha=0.15$ and various values of $g$. The solid line almost following the points for $g=0.05\times\pi/2$ is the fit $P_\textrm{pr}=\exp(-0.0704N^{0.8227}-0.6043)$.}\label{spdkfig}
\end{figure}

The probability of success
\begin{multline}
P_\textrm{pr}=\frac{1}{2^N}\sum_{s_1,\ldots,s_N}
\exp\Bigg\{-\alpha\sum_{n=1}^N\sum_{m=1}^N s_ns_m\\
\times[\cos(g)]^{|m-n|}\cos[2\pi(m-n)/N]\Bigg\}
\end{multline}
is plotted as a function of the number of spins in Fig.~\ref{spdkfig}. $P_\textrm{pr}$ decreases significantly with increasing $|g|$, and for $g=0.05\times\pi/2$, the decrease in $P_\textrm{pr}$ with $N$ is almost exponential as detailed in the figure.

\section{Conclusion}\label{conclusion}

In conclusion, we have proposed and investigated the properties of a specific class of states of a one-dimensional spin chain, and we have suggested a (probabilistic) scheme for preparing the states experimentally, for which the required number of operations scales only polynomially with the total number of spins $N$. The class of states is obtained by replacing the matrices in a matrix product state by chiral vertex operators of a bosonic conformal field theory and truncating these operators to involve only a finite number of bosonic modes. This construction is interesting, because it makes it apparent how one can generate the states by sequential interaction of the spins with an ancilla system, and because the resulting states turn out to have properties that are very different from MPS and also from the critical model obtained without truncating the vertex operators.

Specifically, we have found that the Renyi entropy of a subchain of length $L$ increases significantly, when the vertex operators are truncated, and the functional dependence of the entropy on $L$ turns out to approximately follow a power law in $\sin(\pi L/N)$ for small $M$ as opposed to the logarithmic behavior observed in the critical case. We have thus demonstrated that replacing the finite dimensional matrices in MPS with infinite dimensional operators acting on a single bosonic mode is sufficient to violate the area law and obtain states with high entropy. Considering the $\sigma_z$-$\sigma_z$ correlation function for two spins separated by a distance $k$, we have found undamped oscillatory behavior as a function of $k$ when keeping only a few modes, which is in contrast to the power law decay with $k$ for critical systems and the exponential decay for MPS. These differences mean that the considered class of states provides a framework for describing a particular set of problems, which can not be handled by MPS or critical models. Finally, we have demonstrated how the properties of the states can be altered by modifying the preparation scheme in various ways.

\begin{acknowledgments}
This work has been supported by Carlsbergfondet, the grants FIS2009-11654 and QUITEMAD, and the EU project QUEVADIS.
\end{acknowledgments}

\appendix

\section{Distribution of $\sum_{n=1}^Ls_ne^{2\pi in/N}$}\label{app1}

In this appendix, we determine the probability distribution in the limit of large $N$ for obtaining given values of
\begin{align}
x_L&\equiv\Real\left(\sum_{n=1}^Ls_ne^{2\pi in/N}\right)\label{xL}\\
y_L&\equiv\Imag\left(\sum_{n=1}^Ls_ne^{2\pi in/N}\right)\label{yL}
\end{align}
if we choose random configurations of the spins. For $N$ large, we can to a good approximation replace the angle $\phi\equiv2\pi n/N$ by a continuous parameter and assume that several terms in each of the sums have a value of $\phi$ within an interval of practically infinitesimal width $d\phi$, i.e., $dN=Nd\phi/(2\pi)$ is large compared to one. Considering one such group of $dN$ terms with $\phi\in[\phi_j,\phi_j+d\phi[$, $\phi_j\equiv(j-1)d\phi$, the probability to obtain a given value of $w_j\equiv\sum_{n_j}s_{n_j}$, where $n_j$ labels the terms within the group, is
\begin{equation}
P(w_j)=\frac{1}{2^{dN}}\frac{dN!}{[(dN+w_j)/2]![(dN-w_j)/2]!},
\end{equation}
which in the limit of large $dN$ reduces to a Gaussian distribution
\begin{equation}
P(w_j)=\frac{1}{\sqrt{2\pi dN}}\exp\left(-\frac{w_j^2}{2dN}\right)
\end{equation}
with mean value $0$ and variance $dN$. Rewriting \eqref{xL} and \eqref{yL} into $x_L=\sum_j w_j\cos(\phi_j)$ and $y_L=\sum_j w_j\sin(\phi_j)$, we note that $x_L$ and $y_L$ are linear combinations of random Gaussian variables, and $x_L$ and $y_L$ are hence also random variables with a Gaussian probability distribution $P(x_L,y_L)$. Since $\langle x_L\rangle=\sum_j \langle w_j\rangle\cos(\phi_j)=0$ and $\langle y_L\rangle=\sum_j \langle w_j\rangle\sin(\phi_j)=0$, we conclude that
\begin{equation}
P(x_L,y_L)=\frac{1}{2\pi\sqrt{\det(V)}}\exp\left[-\frac{1}{2}(x_L,y_L)V^{-1}
\left(\begin{array}{c}
x_L\\
y_L
\end{array}\right)\right]
\end{equation}
in the limit of large $N$, where $V$ is the covariance matrix
\begin{equation}
V=\left(\begin{array}{cc}
\langle x_L^2\rangle & \langle x_Ly_L\rangle\\
\langle x_Ly_L\rangle & \langle y_L^2\rangle
\end{array}\right),
\end{equation}
\begin{equation}
\langle x_L^2\rangle=\frac{N}{2\pi}\int_0^{2\pi L/N}\cos^2(\phi)d\phi
=\frac{L}{2}+\frac{N}{8\pi}\sin\left(4\pi\frac{L}{N}\right)
\end{equation}
\begin{equation}
\langle y_L^2\rangle=\frac{N}{2\pi}\int_0^{2\pi L/N}\sin^2(\phi)d\phi
=\frac{L}{2}-\frac{N}{8\pi}\sin\left(4\pi\frac{L}{N}\right)
\end{equation}
\begin{multline}
\langle x_Ly_L\rangle=\frac{N}{2\pi}\int_0^{2\pi L/N}\sin(\phi)\cos(\phi)d\phi\\
=\frac{N}{8\pi}\left[1-\cos\left(4\pi\frac{L}{N}\right)\right]
\end{multline}
and we have used $\langle w_jw_{j'}\rangle=0$ for $j\neq j'$. As a final remark, we note that the probability distribution $\tilde{P}(\tilde{x}_L,\tilde{p}_L)$ of $\tilde{x}_L\equiv\Real\left(\sum_{n=L+1}^Ns_ne^{2\pi in/N}\right)$ and $\tilde{p}_L\equiv\Imag\left(\sum_{n=L+1}^Ns_ne^{2\pi in/N}\right)$ is a Gaussian with $\langle \tilde{x}_L\rangle=\langle \tilde{p}_L\rangle=0$ and covariance matrix $\tilde{V}$ with entries $\langle \tilde{x}_L^2\rangle=N/2-\langle x_L^2\rangle$, $\langle \tilde{p}_L^2\rangle=N/2-\langle p_L^2\rangle$, and $\langle \tilde{x}_L\tilde{p}_L\rangle=-\langle x_Lp_L\rangle$, whereas $\langle x_L\tilde{x}_L\rangle$, $\langle x_L\tilde{p}_L\rangle$, $\langle p_L\tilde{x}_L\rangle$, and $\langle p_L\tilde{p}_L\rangle$ are all zero.

\section{Analytical expression for the Renyi entropy for $M=1$ and $\delta=0$}\label{app2}

Combining \eqref{wave} and \eqref{MCentropi} for $M=1$ and $\delta=0$, we obtain
\begin{multline}\label{eslapp}
e^{-S_L^{(2)}}=\\
\sum_{s_1,\ldots,s_N}\sum_{s'_1,\ldots,s'_N}
e^{\alpha(x_L+iy_L-x'_L-iy'_L)(\tilde{x}_L-i\tilde{y}_L-\tilde{x}'_L+i\tilde{y}'_L)}\\
\times e^{-\alpha[(x_L+\tilde{x}_L)^2+(p_L+\tilde{p}_L)^2 +(x'_L+\tilde{x}'_L)^2+(p'_L+\tilde{p}'_L)^2]}\\
\Bigg/\left(\sum_{s_1,\ldots,s_N} e^{-\alpha[(x_L+\tilde{x}_L)^2+(p_L+\tilde{p}_L)^2]}\right)^2,
\end{multline}
where $x_L$, $p_L$, $\tilde{x}_L$ and $\tilde{p}_L$ are defined in App.~\ref{app1} and $x'_L$, $p'_L$, $\tilde{x}'_L$ and $\tilde{p}'_L$ are the same quantities evaluated for the spin configuration $s'_1,\ldots,s'_N$. For large $N$, the results in App.~\ref{app1} allow us to make the replacement
\begin{multline}
\frac{1}{2^N}\sum_{s_1,\ldots,s_N}\rightarrow\int_{-\infty}^\infty dx_L
\int_{-\infty}^\infty dy_L\int_{-\infty}^\infty d\tilde{x}_L\\
\times\int_{-\infty}^\infty d\tilde{y}_LP(x_L,y_L)\tilde{P}(\tilde{x}_L,\tilde{y}_L)
\end{multline}
and the same for the sum over $s'_1,\ldots,s'_N$. The right hand side of \eqref{eslapp} then turns into a ratio between two Gaussian integrals that are easily evaluated to give Eq.~\eqref{entropy} with
\begin{equation}\label{M1}
M_1=\alpha\left[\begin{array}{cccc}
I_2 & I_2 & 0 & 0\\
I_2 & I_2 & 0 & 0\\
0 & 0 & I_2 & I_2\\
0 & 0 & I_2 & I_2
\end{array}\right],
\end{equation}
\begin{equation}\label{M2}
M_2=-\frac{\alpha}{2}\left[\begin{array}{cccc}
0 & m_2 & 0 & -m_2\\
m_2^T & 0 & -m_2^T & 0\\
0 & -m_2 & 0 & m_2\\
-m_2^T & 0 & m_2^T & 0
\end{array}\right],
\end{equation}
and
\begin{equation}\label{M3}
M_3=\frac{1}{2}\left[\begin{array}{cccc}
V^{-1} & 0 & 0 & 0\\
0 & \tilde{V}^{-1} & 0 & 0\\
0 & 0 & V^{-1} & 0\\
0 & 0 & 0 & \tilde{V}^{-1}
\end{array}\right]
\end{equation}
where $I_2$ is the two-by-two identity matrix, $m_2=I_2+\sigma_y$, and $\sigma_y=\left(\begin{array}{cc}0&-i\\ i&0\end{array}\right)$ is the second of the Pauli matrices.

Repeating the same arguments for the case of nonzero squeezing considered in Sec.~\ref{squeezing}, we again arrive at Eq.~\eqref{entropy}, but the matrices $M_1$ and $M_2$ are replaced by
\begin{multline}
M_1=\alpha\frac{1+4|A|^2}{1-4|A|^2}
\left[\begin{array}{cccc}
I_2 & I_2 & 0 & 0\\
I_2 & I_2 & 0 & 0\\
0 & 0 & I_2 & I_2\\
0 & 0 & I_2 & I_2
\end{array}\right]\\
-\alpha\frac{4|A|}{1-4|A|^2}
\left[\begin{array}{cccc}
\sigma_z & \sigma_z & 0 & 0\\
\sigma_z & \sigma_z & 0 & 0\\
0 & 0 & \sigma_z & \sigma_z\\
0 & 0 & \sigma_z & \sigma_z
\end{array}\right]
\end{multline}
and
\begin{multline}
M_2=-\frac{\alpha}{2}
\left[\begin{array}{cccc}
0 & m_2 & 0 & -m_2\\
m_2^T & 0 & -m_2^T & 0\\
0 & -m_2 & 0 & m_2\\
-m_2^T & 0 & m_2^T & 0
\end{array}\right]\\
-\frac{4\alpha|A|^2}{1-4|A|^2}
\left[\begin{array}{cccc}
0 & I_2 & 0 & -I_2\\
I_2 & 0 & -I_2 & 0\\
0 & -I_2 & 0 & I_2\\
-I_2 & 0 & I_2 & 0
\end{array}\right]\\
+\frac{2\alpha|A|}{1-4|A|^2}
\left[\begin{array}{cccc}
0 & \sigma_z & 0 & -\sigma_z\\
\sigma_z & 0 & -\sigma_z & 0\\
0 & -\sigma_z & 0 & \sigma_z\\
-\sigma_z & 0 & \sigma_z & 0
\end{array}\right],
\end{multline}
where $\sigma_z=\left(\begin{array}{cc}1&0\\0&-1\end{array}\right)$ is the third of the Pauli matrices.

\section{Approximate independence of $\sum_{n=1}^Ns_ne^{2\pi inp/N}$ and $\sum_{n=1}^Ns_ne^{2\pi inq/N}$ for $p\neq q$}\label{app3}

Assuming we can drag the Gaussian approximation in App.~\ref{app1} to the limit, where each individual spin is replaced by a Gaussian random variable $w_n$ with mean $0$ and variance $1$, we can write the real and imaginary parts $x_N^{(p)}$ and $y_N^{(p)}$ of $\sum_{n=1}^Ns_ne^{2\pi inp/N}$, where $p$ is a positive integer, as
\begin{align}
x_N^{(p)}&=\sum_{n=1}^Nw_n\cos(2\pi np/N),\\
p_N^{(p)}&=\sum_{n=1}^Nw_n\sin(2\pi np/N),
\end{align}
and it follows that
\begin{align}
\langle x_N^{(p)}x_N^{(q)}\rangle&=\frac{N}{2\pi}\int_0^{2\pi}
\cos(p\phi)\cos(q\phi)d\phi=\delta_{pq}N/2,\\
\langle p_N^{(p)}p_N^{(q)}\rangle&=\frac{N}{2\pi}\int_0^{2\pi}
\sin(p\phi)\sin(q\phi)d\phi=\delta_{pq}N/2,\\
\langle x_N^{(p)}p_N^{(q)}\rangle&=\frac{N}{2\pi}\int_0^{2\pi}
\cos(p\phi)\sin(q\phi)d\phi=0.
\end{align}
For a Gaussian probability distribution, this is sufficient to conclude that $x_N^{(p)}$ and $p_N^{(p)}$ are independent of $x_N^{(q)}$ and $p_N^{(q)}$ for $p\neq q$. We hence expect $\sum_{n=1}^Ns_ne^{2\pi inp/N}$ and $\sum_{n=1}^Ns_ne^{2\pi inq/N}$ to be approximately independent for $p\neq q$.

Another way to think of the independence of $\sum_{n=1}^Ns_ne^{2\pi inp/N}$ and $\sum_{n=1}^Ns_ne^{2\pi inq/N}$ is to regard the terms in the sums as points positioned at $e^{2\pi inp/N}$ ($e^{2\pi inq/N}$) in the complex plane and associated with either a minus or a plus depending on the value of $s_n$. One can divide these points into groups with similar values of $2\pi np/N$ ($2\pi nq/N$) modulus $2\pi$, and the value of $\sum_{n=1}^Ns_ne^{2\pi inp/N}$ ($\sum_{n=1}^Ns_ne^{2\pi inq/N}$) is determined by the number of points associated with plus and with minus within each group. By changing the phase of the points from $2\pi np/N$ to $2\pi nq/N$, a reorganization of the division of the points into groups occurs, which depends on the precise value of $2\pi np/N$ for the points associated with plus or minus belonging to the same group before the change. When there are many points within each group, it seems plausible that the values of $\sum_{n=1}^Ns_ne^{2\pi inq/N}$ attainable for a fixed value of $\sum_{n=1}^Ns_ne^{2\pi inp/N}$ are almost independent of the value of $\sum_{n=1}^Ns_ne^{2\pi inp/N}$, except for very special choices of $\sum_{n=1}^Ns_ne^{2\pi inp/N}$, which justifies the assumption of independence. If, however, we consider several sums of the form $\sum_{n=1}^Ns_ne^{2\pi inp/N}$ with different values of $p$, more constraints are present for the possible values of one of them for all the others fixed. Hence, we expect the assumption of independence to break down if too many sums of the form $\sum_{n=1}^Ns_ne^{2\pi inp/N}$ are involved.

\section{Analytical expression for the $\langle\sigma_z(0)\sigma_z(k)\rangle$ correlation function}\label{app4}

In this appendix, we provide analytical arguments for the approximate expressions \eqref{corM} and \eqref{sqacor} for the $\langle\sigma_z(0)\sigma_z(k)\rangle$ correlation function. We first consider the case without squeezing. Since $\langle\sigma_z(0)\sigma_z(k)\rangle=\langle\sigma_z(n)\sigma_z(n+k)\rangle$, $n=1,2,\ldots,N$, it follows from \eqref{wave} and \eqref{correlator} that
\begin{widetext}
\begin{equation}\label{correlatorapp}
\langle\sigma_z(0)\sigma_z(k)\rangle=
\frac{\sum_{s_1,\ldots,s_N}(1-\delta+\delta\delta_s)\sum_{p=1}^Ns_ps_{p+k}
\exp\left[-\alpha\sum_{n=1}^N\sum_{m=1}^N s_ns_{n+m}\sum_{q=1}^M\frac{1}{q}\cos(2q\pi m/N)\right]} {\sum_{s_1,\ldots,s_N}(1-\delta+\delta\delta_s)
\exp\left[-\alpha\sum_{n=1}^N\sum_{m=1}^N s_ns_{n+m}\sum_{q=1}^M\frac{1}{q}\cos(2q\pi m/N)\right]N}.
\end{equation}
\end{widetext}
This expression can be simplified by noting that
\begin{multline}
\sum_{n=1}^Ns_ns_{n+m}=N\sum_{q=1}^Ne^{2\pi imq/N}|\tilde{s}_q|^2=
N\Bigg[|\tilde{s}_N|^2\\
+(-1)^m|\tilde{s}_{N/2}|^2+2\sum_{q=1}^{N/2-1}\cos(2\pi mq/N)|\tilde{s}_q|^2\Bigg],
\end{multline}
\begin{equation}
\sum_{m=1}^N\sum_{n=1}^Ns_ns_{n+m}\sum_{q=1}^M\frac{1}{q}\cos(2\pi qm/N)
=N^2\sum_{q=1}^M\frac{1}{q}|\tilde{s}_q|^2,
\end{equation}
and (for $k\neq N$ modulus $N$)
\begin{equation}
1+2\sum_{p=1}^{N/2-1}\cos(2\pi kp/N)+(-1)^k=0,
\end{equation}
where $\tilde{s}_q$ is the discrete Fourier transform of $s_n$ defined in Eq.~\eqref{tildesq}. For $k=N$ modulus $N$, \eqref{correlatorapp} reduces to $\langle\sigma_z(0)\sigma_z(N)\rangle=1$, and for all other values of $k$,
\begin{multline}\label{cor2}
\langle\sigma_z(0)\sigma_z(k)\rangle=F_M(N)-F_M(N/2)\\
-2\sum_{p=1}^{N/2-1}\cos(2\pi kp/N)\left[F_M(N/2)-F_M(p)\right],
\end{multline}
where
\begin{equation}
F_M(p)=\frac{\sum_{\textrm{conf}}|\tilde{s}_p|^2
\exp\left(-\alpha N^2\sum_{q=1}^M\frac{1}{q}|\tilde{s}_q|^2\right)} {\sum_{\textrm{conf}}\exp\left(-\alpha N^2\sum_{q=1}^M\frac{1}{q}|\tilde{s}_q|^2\right)}
\end{equation}
is the expectation value of $|\tilde{s}_p|^2$ and $\sum_{\textrm{conf}}$ is the sum over all allowed configurations of the spins. Note, in particular, that the constraint $\sum_{n=1}^Ns_n=0$ for $\delta=1$ translates into $\tilde{s}_N=0$, whereas
\begin{equation}\label{ssqs}
\sum_{n=1}^Ns_n^2=N \quad\Leftrightarrow\quad \sum_{q=1}^N|\tilde{s}_q|^2=1.
\end{equation}

If we assume $\tilde{s}_p$ and $\tilde{s}_q$ to be independent as discussed in App.~\ref{app3}, we can write
\begin{equation}\label{FM}
F_M(p)\approx\left\{\begin{array}{l}
\frac{\sum_{\textrm{conf}}|\tilde{s}_p|^2
\exp\left(-\alpha N^2|\tilde{s}_p|^2/p\right)} {\sum_{\textrm{conf}}\exp\left(-\alpha N^2|\tilde{s}_p|^2/p\right)}
\textrm{ for }1\leq p\leq M\\
\frac{\sum_{\textrm{conf}}|\tilde{s}_p|^2} {\sum_{\textrm{conf}}1}
\textrm{ for }M<p\leq N/2\textrm{ and }p=N
\end{array}
\right..
\end{equation}
The sum
\begin{equation}\label{sqs}
\sum_{\textrm{conf}}|\tilde{s}_p|^2=
\frac{1}{N^2}\sum_{n=1}^N\sum_{m=1}^Ne^{2\pi i(m-n)p/N}\sum_{\textrm{conf}}s_ns_m
\end{equation}
can be calculated exactly. For the case, where the zero mode is included, the total number of spin configurations is $N!/[(N/2)!]^2$, the number of spin configurations with $s_ns_m=+1$ is $2(N-2)!/[(N/2-2)!(N/2)!]$ for $n\neq m$, the number of spin configurations with $s_ns_m=-1$ is $2(N-2)!/[(N/2-1)!(N/2-1)!]$ for $n\neq m$, and consequently
\begin{equation}
\sum_{\textrm{conf}}s_ns_m=\left\{\begin{array}{rl}
\frac{N!}{((N/2)!)^2}&\textrm{for }n=m\\
\frac{-1}{N-1}\frac{N!}{((N/2)!)^2}&\textrm{for }n\neq m\end{array}\right..
\end{equation}
For the case, where the zero mode is not included, the total number of spin configurations is $2^N$, the number of spin configurations with $s_ns_m=+1$ is $2^{N-1}$ for $n\neq m$, the number of spin configurations with $s_ns_m=-1$ is $2^{N-1}$ for $n\neq m$, and
\begin{equation}
\sum_{\textrm{conf}}s_ns_m=\left\{\begin{array}{cl}
2^N&\textrm{for }n=m\\
0&\textrm{for }n\neq m\end{array}\right..
\end{equation}
When we insert this into \eqref{sqs}, we obtain
\begin{equation}\label{sqsN}
\frac{\sum_{\textrm{conf}}|\tilde{s}_p|^2}
{\sum_{\textrm{conf}}1}=\left\{\begin{array}{cl}
1/(N-\delta)&\textrm{for }p=1,2,\ldots,N-1\\
(1-\delta)/N&\textrm{for }p=N
\end{array}\right..
\end{equation}
Note that this result is consistent with \eqref{ssqs}.

Returning to the expression \eqref{FM} for $F_M(p)$ for $1\leq p\leq M$, we note that the exponential is small unless $|\tilde{s}_p|^2$ is of order $p/(\alpha N^2)$ or smaller. The value of $F_M(p)$ for $1\leq p\leq M$ is hence at least about a factor of $N$ smaller than the value of $F_M(p)$ for $M<p\leq N/2$. In conclusion, we thus have
\begin{equation}\label{res1}
F_M(p)\approx\left\{\begin{array}{cl}
0 & \textrm{for }1\leq p\leq M\\
1/N & \textrm{for }M+1 \leq p \leq N/2\\
(1-\delta)/N & \textrm{for }p=N
\end{array}\right.,
\end{equation}
which, when combined with \eqref{cor2}, leads to \eqref{corM}. We note that \eqref{res1} is, in fact, inconsistent with \eqref{ssqs}, and the inconsistency grows with $M$. This indicates that the approximation is less accurate for higher $M$ as expected.

For the case of nonzero squeezing investigated in Sec.~\ref{squeezing}, the derivation follows similar steps. In particular, $F_1(p)$ is still given by \eqref{res1} for $p\neq1$. For $p=1$, however, we can no longer assume that $F_1(1)$, which in this case is given by
\begin{multline}
F_1(1)=\\
\frac{\sum_{s_1,\ldots,s_N}(x_N^2+y_N^2)\exp\left(
-\alpha\frac{1-2|A|}{1+2|A|}x_N^2
-\alpha\frac{1+2|A|}{1-2|A|}y_N^2\right)}
{N^2\sum_{s_1,\ldots,s_N}\exp\left(
-\alpha\frac{1-2|A|}{1+2|A|}x_N^2
-\alpha\frac{1+2|A|}{1-2|A|}y_N^2\right)},
\end{multline}
is approximately zero ($x_N$ and $y_N$ are defined in \eqref{xL} and \eqref{yL}, respectively). In fact, for $|A|\rightarrow(1/2)^-$, only configurations with $y_N=0$ contribute, and for these configurations the exponential factor is unity, while $x_N$ may take different values. Instead, we evaluate the expectation value by utilizing the probability distribution for $x_N$ and $p_N$ derived in App.~\ref{app1}, and this leads to \eqref{sqacor}.

\end{document}